\documentclass[preprint,12pt]{elsarticle}

\usepackage{amssymb}

\usepackage[T1]{fontenc}
\usepackage{natbib}
\usepackage[hidelinks]{hyperref}
\usepackage{caption}
\usepackage{url}

\journal{Results in engineering}

\begin{document}

\begin{frontmatter}




\title{The Sustainable Development Goals and Aerospace Engineering: A critical note through Artificial Intelligence}

\author[label1]{Alejandro S\'anchez-Roncero}
\author[label1,label6]{\`Oscar Garibo-i-Orts} 
\author[label1]{J. Alberto Conejero}
\author[label2]{Hamidreza Eivazi}
\author[label2]{Ferm\'in Mallor}
\author[label2]{Emelie Rosenberg}
\author[label3,label4]{Francesco Fuso-Nerini}
\author[label5]{Javier Garc\'ia-Mart\'inez}
\author[label2,label4]{Ricardo Vinuesa}
\author[label1]{Sergio Hoyas\corref{cor1}}

\cortext[cor1]{Corresponding author, serhocal@mot.upv.es}
\address[label1]{Instituto de Matem\'atica Pura y Aplicada, Universitat Polit\`ecnica de Val\`encia, Camino de Vera 46024 Val\`encia, Spain}
\address[label2]{ACES -- Association of Spanish Scientists in Sweden, Stockholm, Sweden}
\address[label3]{Division of Energy Systems, KTH Royal Institute of Technology, Stockholm, Sweden} 
\address[label4]{KTH Climate Action Centre, Stockholm, Sweden}
\address[label5]{Molecular Nanotechnology Lab, Department of Inorganic Chemistry, University of Alicante,Alicante, Spain}
\address[label6]{GRID - Grupo de Investigación en Ciencia de Datos\\ Valencian International University - VIU,Val\`encia, Spain}

\begin{abstract}
The 2030 Agenda of the United Nations (UN) revolves around the Sustainable Development Goals (SDGs). A critical step towards that objective is identifying whether scientific production aligns with the SDGs' achievement. To assess this, funders and research managers need to manually estimate the impact of their funding agenda on the SDGs, focusing on accuracy, scalability, and objectiveness. With this objective in mind, in this work, we develop ASDG, an easy-to-use artificial-intelligence (AI)-based model for automatically identifying the potential impact of scientific papers on the UN SDGs. As a demonstrator of ASDG, we analyze the alignment of recent aerospace publications with the SDGs. The Aerospace data set analyzed in this paper consists of approximately 820,000 papers published in English from 2011 to 2020 and indexed in the Scopus database. The most-contributed SDGs are 7 (on clean energy), 9 (on industry), 11 (on sustainable cities) and 13 (on climate action). The establishment of the SDGs by the UN in the middle of the 2010 decade did not significantly affect the data. However, we find clear discrepancies among countries, likely indicative of different priorities. Also, different trends can be seen in the most and least cited papers, with clear differences in some SDGs. Finally, the number of abstracts the code cannot identify is decreasing with time, possibly showing the scientific community's awareness of SDG.  
\end{abstract}

\begin{graphicalabstract}
\includegraphics[scale=0.55]{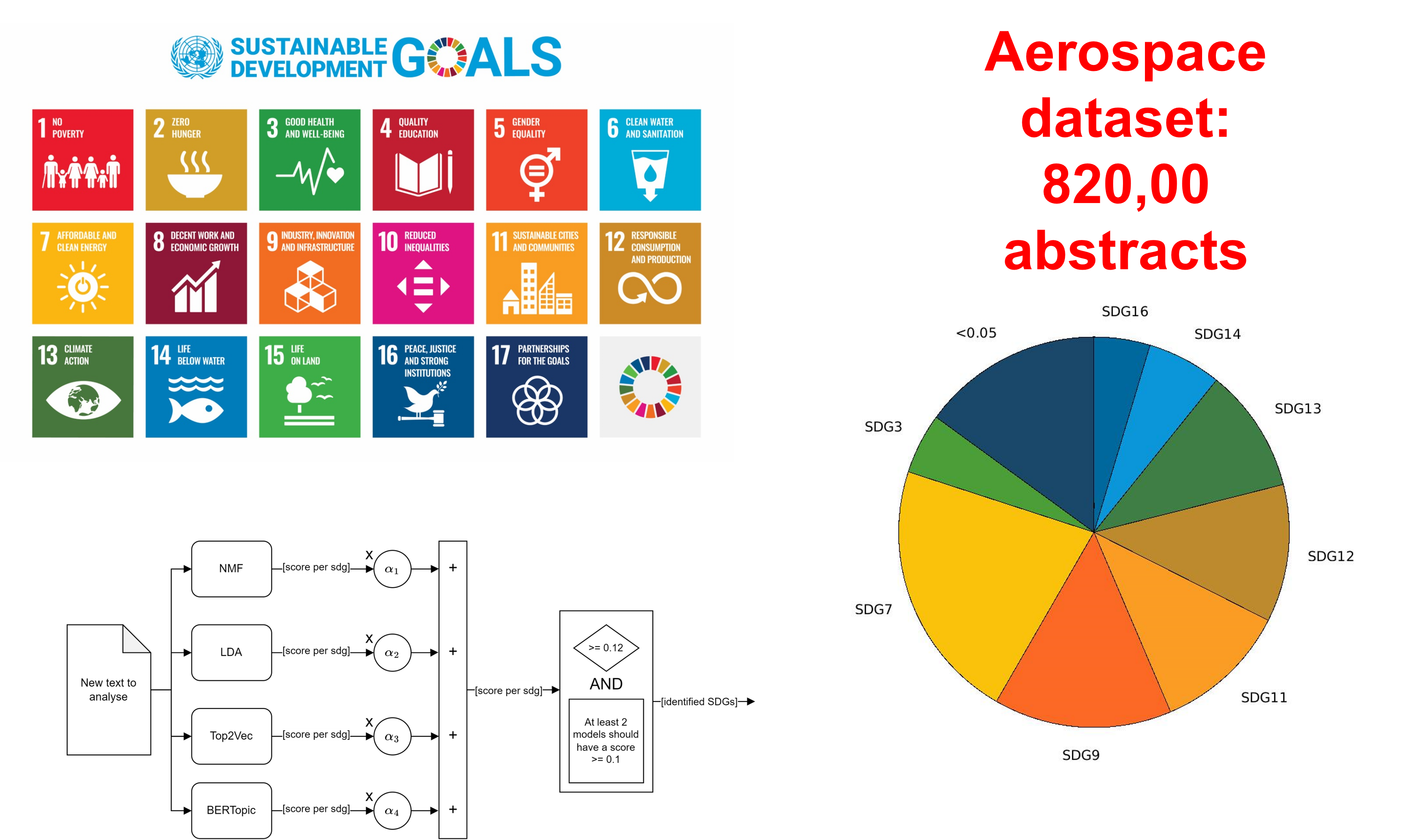}
\end{graphicalabstract}

\begin{highlights}
\item Analysis of the alignment of Aerospace Engineering to the Sustainable Development Goals
\item Aerospace Engineering community seems to be aligned with SDG7, SDG9, SDG12 and SDG13.
\end{highlights}


\end{frontmatter}


\section{Introduction}
In 2015 all state members of the United Nations (UN) adopted the 2030 Agenda for Sustainable Development. The UN intends to promote peace and prosperity for people and the planet with a vision for the near future. To make that vision a reality, the 2030 Agenda consists of 17 Sustainable Development Goals (SDGs) \cite{UNGA}. They represent the actions that countries from all over the world (both developed and developing) should implement as global cooperation for the future of our planet.

The 17 SDGs, see description in \cite{UNGA}, are as follows (those most closely related to Aerospace Engineering have been written in italic font): 
\begin{itemize}
\item {\includegraphics[scale=0.15]{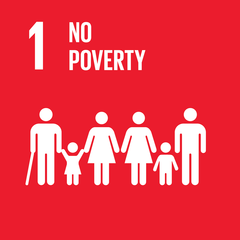}}
SDG 1: End poverty in all its forms everywhere.

\item {\includegraphics[scale=0.15]{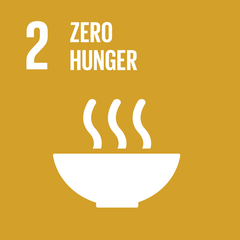}} SDG 2: End hunger, achieve food security and improved nutrition and promote sustainable agriculture.

\item {\includegraphics[scale=0.15]{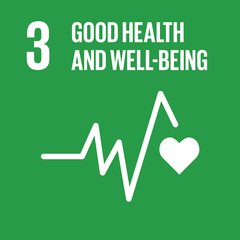}} SDG 3: Ensure healthy lives and promote well-being for all at all ages.

\item {\includegraphics[scale=0.15]{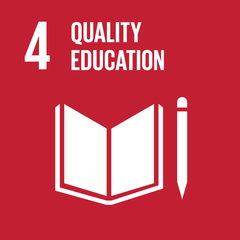}} SDG 4: Ensure inclusive and equitable quality education and promote lifelong learning opportunities for all.

\item {\includegraphics[scale=0.15]{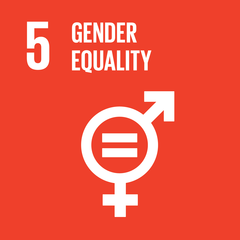}} SDG 5: Achieve gender equality and empower all women and girls.

\item {\includegraphics[scale=0.15]{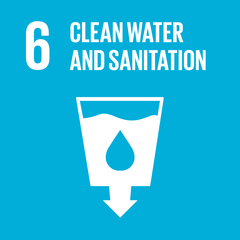}} SDG 6: Ensure availability and sustainable management of water and sanitation for all.

\item {\includegraphics[scale=0.15]{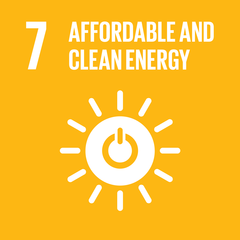}} SDG 7: {\it Ensure access to affordable, reliable, sustainable, and modern energy for all}.

\item {\includegraphics[scale=0.15]{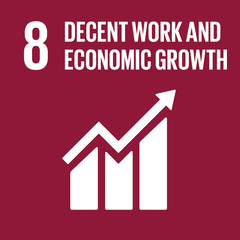}} SDG 8: Promote sustained, inclusive, and sustainable economic growth, full and productive employment, and decent work for all.

\item {\includegraphics[scale=0.15]{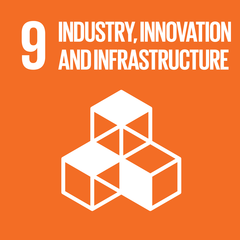}} SDG 9: {\it Build resilient infrastructure, promote inclusive and sustainable industrialization, and foster innovation}.

\item {\includegraphics[scale=0.15]{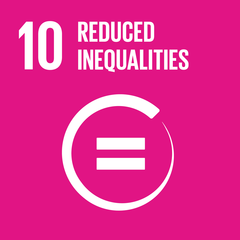}} SDG 10: Reduce inequality within and among countries.

\item {\includegraphics[scale=0.15]{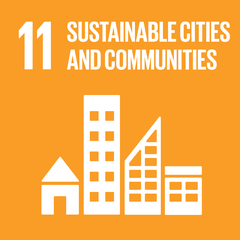}} SDG 11: {\it Make cities and human settlements inclusive, safe, resilient and sustainable}.

\item {\includegraphics[scale=0.15]{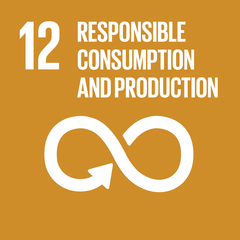}} SDG 12: {\it Ensure sustainable consumption and production patterns}.

\item {\includegraphics[scale=0.15]{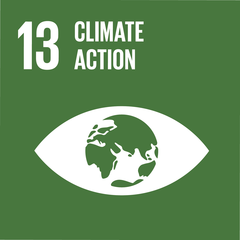}} SDG 13: {\it Take urgent action to combat climate change and its impacts}.

\item {\includegraphics[scale=0.15]{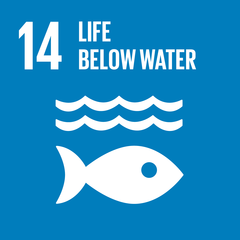}} SDG 14: Conserve and sustainably use the oceans, seas, and marine resources for sustainable development.

\item {\includegraphics[scale=0.15]{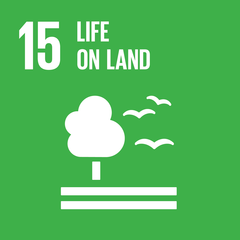}} SDG 15: Protect, restore and promote sustainable use of terrestrial ecosystems, sustainably manage forests, combat desertification, and halt and reverse land degradation and halt biodiversity loss.

\item {\includegraphics[scale=0.15]{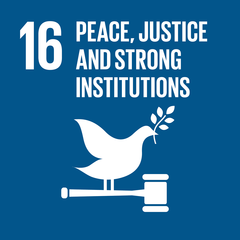}} SDG 16: Promote peaceful and inclusive societies for sustainable development, provide access to justice for all and build effective, accountable and inclusive institutions at all levels.

\item {\includegraphics[scale=0.15]{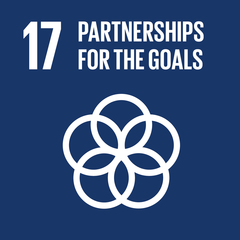}} SDG 17: Strengthen the means of implementation and revitalize the Global Partnership for Sustainable Development.
\end{itemize}



It is important to note that the recent paradigm change introduced by the fast digitalization of business, academics, daily life, and even policy-making is profound. A recent study by Vinuesa et al. \cite{Vinuesa2020} examined in-depth how artificial intelligence (AI) affects the accomplishment of the UN's 2030 Agenda. Although they discovered that $79\%$ of the aims would be positively affected by AI, they also noted that the growth of AI could hinder or even have a detrimental impact on the achievement of $35\%$ of these targets. The SDGs are all interconnected, and while there are numerous synergies, it is vital to recognize and properly document any trade-offs to reach the full potential of AI's ability to contribute to creating a sustainable future.  Furthermore, Gupta et al. \cite{gupta2021assessing} extended their work to discussions on the implications of AI on the SDGs at the indicator level. In this regard, it is crucial to emphasize that implementing clear and understandable strategies requires employing
AI-based technologies to achieve the SDGs. According to Vinuesa and Sirmaeck \cite{vinuesa_natmi}, deploying interpretable AI would produce an algorithmic usage that focuses on accountability and transparency. This has important implications in the context of AI regulations, particularly when focusing on public-policy applications towards achieving the SDGs \cite{goh2021regulating}.

The UN SDGs, must be accomplished while we are amid a climate emergency, as confirmed in the last Intergovernmental Panel on Climate Change \cite{IPC21}. This is particularly important in the case of Aerospace Engineering \cite{francesco_energy,francesco_climate}. To summarize the importance of aerodynamics, for example, about a quarter of today's energy is spent moving fluids along pipes or vehicles through air or water. Turbulence dissipates $25\%$ of this energy, which is responsible for up to $5\%$ of the CO2 dumped by humanity every year\cite{jim13}. Considering that 340 billion liters of fuel were used in 2017 for air transportation worldwide (as reported by IATA \cite{IATA}), there is considerable potential for energy savings and fuel consumption reduction. Before the coronavirus-disease-19 (COVID-19) pandemic, this quantity was growing at a 3\% rate yearly. It is expected to return to this level once the pandemic is over \cite{IATA}. 

Furthermore, since the SDGs are interconnected, it is not always easy to determine the appropriate policies for achieving a particular goal without substantially jeopardizing the accomplishment of others. This effect is shown in the thorough analysis by Pham-Truffert et al. ~\cite{Pham_Truffert}. They collated and described the various good and negative implications when attempting to accomplish all of the SDGs. It is important to note that, despite being valuable, this type of study is frequently skewed toward the opinions of the specialists participating in the process. Occasionally, the positive and bad correlations are relatively straightforward. SDGs 1 (no poverty) and 3 (good health) will both be significantly impacted by efforts to address SDG 2 (no hunger), for example. Thus, it is crucial to attempt to get past these obvious findings and investigate more ambitious methods to uncover less apparent links, reference ISC report. This theory suggests being able to evaluate a sizable number of literature references while reducing biases. We believe the best way to achieve this goal is to use state-of-the-art data-driven methodologies.

With this in mind, a preliminary version of our code ASDG (Automatic Classification of Impact to Sustainable Development Goals) can be found in \cite{san22}. We believe  that a promising way to achieve significant progress in the SDG Agenda is by using AI-based methods to inform policy decisions to maximize the synergies and minimize the trade-offs. With this goal in mind, we created ASDG. This AI-based framework constitutes a step in this direction by enabling the automatic classification of hundreds of thousands of scientific papers by their impact on each SDG. As such, ASDG can be seen as a contribution to SDG17. This work presents the first application of ASDG: aerospace engineering. A brief summary of ASDG is given in the next section, together with a description of the database. The results are explained in the third section. Conclusions and future work are described in the last section. 
\begin{figure}[h]
  \centering
  \includegraphics[width=\linewidth]{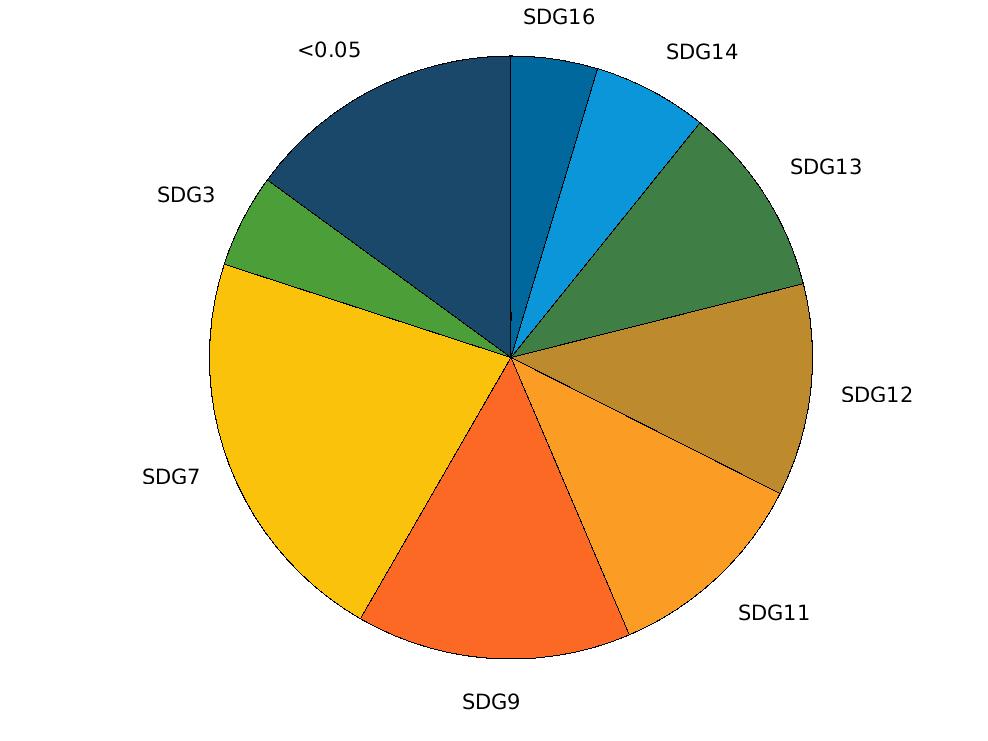}
  \caption{Distribution of the database based on the corresponding SDGs. Note that the SDGs that are not explicitly present here are subsumed in the $``\le 0.05"$ portion}
  \label{fig:figure1}
\end{figure}
\section{Methods}

The code employed for this article, ASDG, can identify the connection between a paper and an SDG through its abstract. It uses four different models: Non-Negative Matrix Factorization (NMF)~\cite{lee1999learning}, Distributed Representations of Topics (Top2Vec)~\cite{angelov20},  Latent Dirichlet Allocation (LDA) ~\cite{blei2003latent}, and BERTopic \cite{ref:bertopic}. Due to their inherently different nature, the information that each model extracts from a text is different. In other words, their functionalities are complementary. To take advantage of this fact, ASDG introduces a voting mechanism. Similar ideas have been used very recently for studying the social network Twitter \cite{ref:comparison_all}. In the voting stage, ASDG takes the scores of each model for each text as inputs. Using this information, ASDG decides which identified SDGs have enough confidence to assume that the text relates to them.

The validation of ASDG was carried out in a previous publication \cite{san22}. The model's training (based on 510 manually-curated text files related to each SDG) was described in that work. Briefly, after downloading all papers referenced in \cite{Vinuesa2020}, for a total of 186  works, we manually selected papers with at least an Abstract and Body differentiated, extracting the sections manually in 40\% of them. A Deep Neural Network \cite{GROBID} was used to extract the remaining $60\%$ automatically. This tool is based on images instead of converting the pdf file to text. We validated this tool with the manually extracted pdf files and checked out every abstract. As the authors of \cite{Vinuesa2020} classified all these papers based on an expert consensus, we labelled these papers to classify all these papers correctly, obtaining an $80\%$ agreement.

The methods mentioned above are briefly described next:

\subsection {NMF}
Non-negative Matrix Factorization model (NMF) \cite{lee1999learning}. This method can reduce the space dimension of the problem, extracting essential features. We consider 16 topics, as SDG 17 is currently not considered. All training and validation texts have been preprocessed. This includes removing accents, adding lemmatization, and allowing bigrams. 

\subsection{Top2Vec}
A Top2Vec model \cite{angelov20} was trained using the embedding model "all-MiniLM-L6-v2". This embedding was pre-trained on a larger corpus, which works better when the training corpus is small. Preprocessing is not required, but accents and non-numeric characters still need to be removed.
In this case, no document segmentation is defined. The extraction of topics was unsupervised. Since the association of the training texts with the SDGs was known beforehand, we queried the associated texts and their scores for each topic, creating an association matrix as it was done with the NMF model.

\subsection{LDA}
A Latent Dirichlet Allocation model \cite{blei2003latent} was also trained with the following configuration:
\begin{itemize}
    \item Number of topics: 16.
    \item Passes: 400. Iterations: 1000. Chunk size: 2000
    \item Bigrams are allowed
    \item Minimum word count: 1, Maximum word frequency: 0.7
\end{itemize}
The training and validation texts were preprocessed similarly to the NMF case. In this case, the model assumes that the documents follow a Dirichlet distribution over topics, and topics over words. Thus, it inherently allows having more than 1 topic in each document. The association matrix was calculated as with the other models. Only the UN training texts were used, no extra files. Note that this method has been successfully employed to automatically classify the AI curricula of a wide range of universities based on their respective contents ~\cite{javed2022get}.

\subsection{Bertopic}
BERTopic is a topic modeling technique very similar to Top2Vec  since both are unsupervised clustering-based techniques \cite{ref:bertopic}. BERTopic extracts coherent topic representation via implementing a class-based variation of the term frequency-inverse document frequency (TF-IDF). The steps it follows are:

\begin{enumerate}
\item Generating the document embeddings with a pretrained transformer-based language model. The embedded words which are semantically similar will be placed close to each other in semantic space. In this way, document-level information is extracted from the corpora. 

\item The document embeddings are dimensionally reduced. This is because as data increases dimensionality, the distance to the closest point tends to approach the distance to the farthest point. As a result, in high dimensional space, spatial locality becomes ill-defined, and distance measures differ little \cite{ref:bertopic}.

\item A density-based method clusters are created. This technique assumes that words near the cluster's centroid are most representative of that cluster. However, in practice, a cluster will not always lie within a sphere around a cluster centroid which might conduce to the extraction of misleading topics.

\item Topics vectors are extracted from the cluster. A class-based version of TF-IDF is used to overcome the limitation of the centroid-based perspective. This has the advantage of separating the clustering technique from the topic generation, allowing more flexibility.
\end{enumerate}

\subsection{Voting}

A combination of the previously described model is used to take advantage of their respective strengths, as the models complement each other. After a careful study, one document is linked to an SDG if:
\begin{itemize}
    \item Any model's score on an SDG is greater than $0.4$ (maximum 0.5), or
    \item The model's score on an SDG is greater than $0.1$ for LDA and BerTopic. 
\end{itemize}
Using this voting system, we successfully classified $81\%$ of the papers based only on the information in the abstract. 

Regarding the database, we have downloaded 820,000 documents, comprising articles and conference papers, from Scopus database \cite{scopus}. The search criterion relied on seeking the words "aerospace," "aeronautics," "aeronautical," and "aviation" in all the metadata of the papers. For obvious reasons, the language of the document must be English. This procedure may lead to over-represented affiliations in English, and some papers, i.e., \cite{hoy22,obe22}, are not found based on these keywords. However, the casuistic can be extremely long, and it is impossible to add every possible author to the list. Nevertheless, the number of papers studied is high. We firmly believe it is representative of the state of Aerospace Engineering to SDGs, as we are analysing a production of more than 80,000 papers yearly. 

\section{Results}
\begin{figure}[h]
  \centering
  \includegraphics[width=\linewidth]{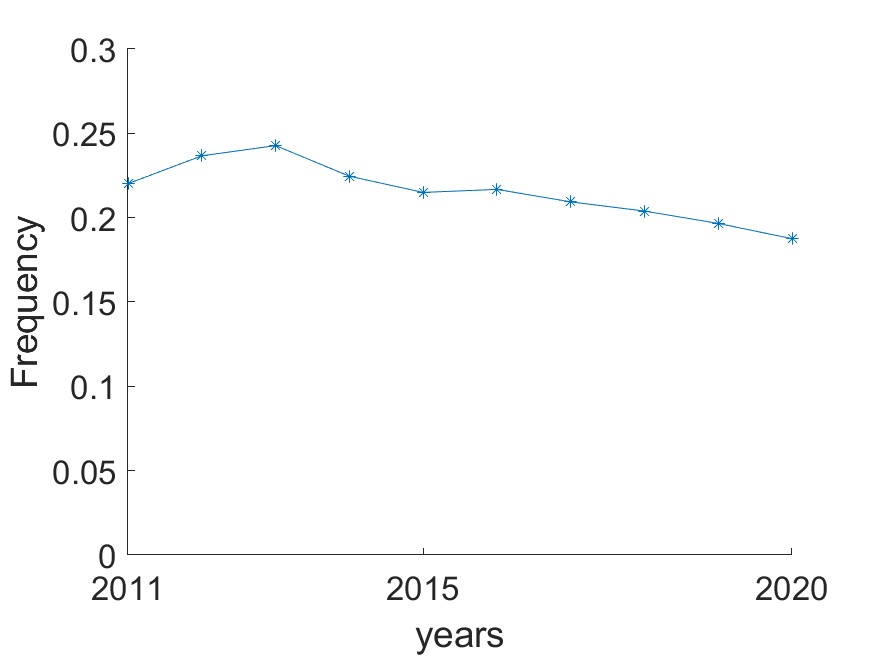}
  \caption{Distribution in the frequency of not identified papers by year from 2010 (left) to 2020(right). The frequency is obtained by dividing the papers not assigned over the total papers of that year}
  \label{fig:figure2n}
\end{figure}

ASDG was used to analyze the aforementioned database with the configuration described above. Out of approximately 167,000 abstracts, 21\% were not assigned any SDG and thus were removed from the discussion. However, there is a positive point in these papers. As shown in figure \ref{fig:figure2n}, the number of documents not identified is decreasing. This could mean that the authors are aware of the existence of the SDGs and are adapting their abstracts. 

The global situation can be seen in figure \ref{fig:figure1}. A 75\% of the works are related to SDGs 7,  9, 11, 12, 13, 14, and 16. SDGs 1, 2, 5, 8, and 10 are below 0.01 in frequency. Here we have defined frequency as the number of papers associated with an SDG over the total number of documents. In the next figures, we will use a similar frequency definition but yearly. As the number of documents increases steadily by at least $6\%$ yearly, we think the frequency is a better parameter than the number of papers.

\begin{figure}[h]
  \centering
  \includegraphics[width=\linewidth]{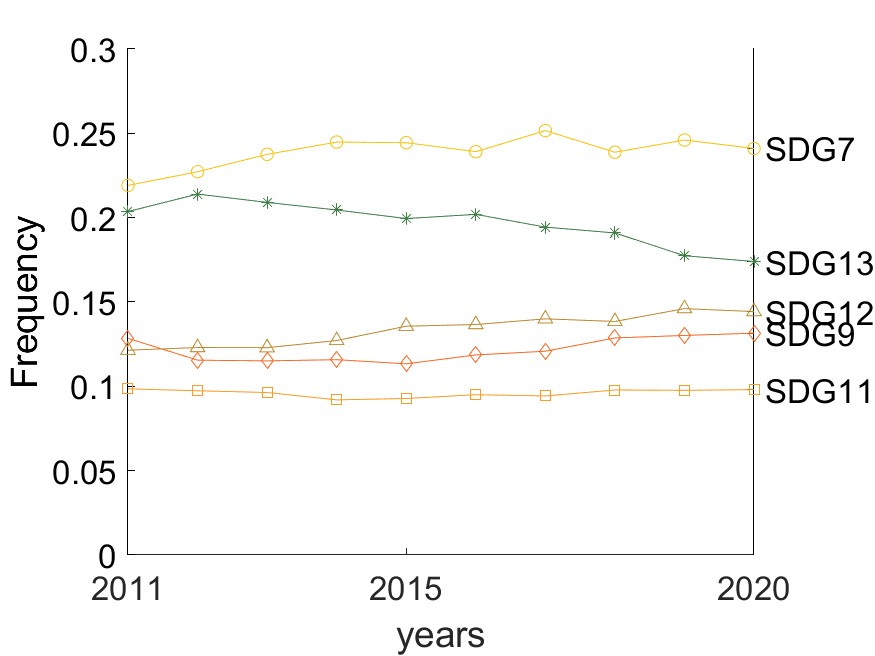}
  \caption{Distribution in the frequency of every SDGs by year from 2010 (left) to 2020(right). For every SDG and year, the frequency is obtained by dividing the papers assigned to a particular SDG over the total papers of that year}
  \label{fig:figure2}
\end{figure}

\begin{figure}[h]
  \centering
  \includegraphics[width=0.95\linewidth]{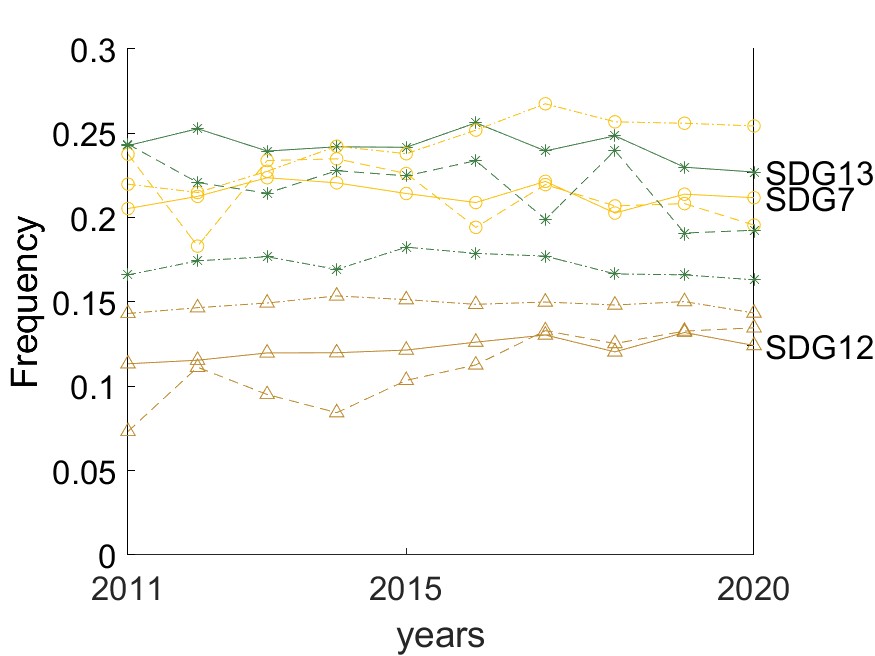}
  
  \caption{Distribution of SDGs 7, 12, and 13 by year for articles from authors in the United States (continuous line), China (dash-dot), and Spain (dash-dash). Symbols: SDG7 circles, SDG12 triangles, and SDG13 stasrs}
  \label{fig:figure3}
\end{figure}

\begin{figure}[h]
  \centering
  \includegraphics[width=0.95\linewidth]{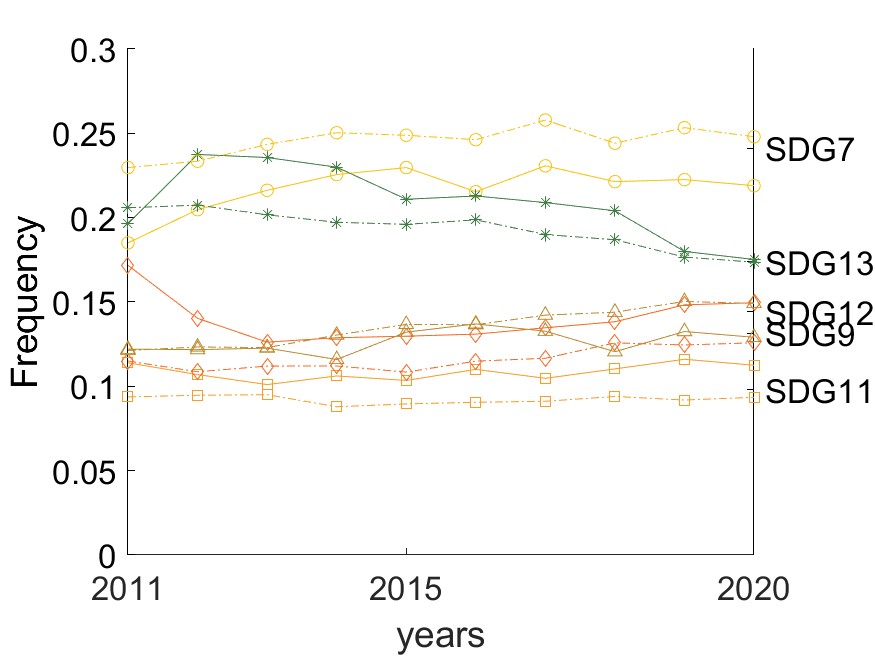}
  \caption{Evolution of the most (dash-dot) and less (continuous) cited papers. Symbols as in figure 2}
  \label{fig:figure5}
\end{figure}

One of the most important points is identifying whether researchers are steering their work toward fulfilling the SDGs. As the SDGs were launched in 2015, there has been enough time to see their introduction's consequences. Figure \ref{fig:figure2} shows the trends with years of the higher-frequency SDGs. The leftmost point is the year 2011, advancing to 2020. Not a clear pattern emerges from this figure. The trends in the most popular SDGs seem to be present before the appearance of SDGs. While SDG 7 continues with a constant value of close to 0.24 in the last 7 years, SDG 12 and 13 reduce their distance. This means that climate action is getting reduced as responsible production grows. As both SDGs are tightly coupled, this is probably not as serious as it seems from the point of view of climate action.

ASDG can also be used to analyze the priorities of different countries. In Figure \ref{fig:figure3} we can see the same study for the USA, China, and Spain. Since 10 years ago Aerospace Engineering community in the USA has been focused on climate action. However, China seems to focus on engineering and production instead of climate action. A small country like Spain seems to be in the middle, with larger values in other SDGs not present in this plot. Again, the introduction of the SDGs does not seem to change the patterns. 

Finally, we centre our analysis on whether highly-cited works exhibit the same trend as the quorum. In Figure \ref{fig:figure5}, we show the results for the most and the least cited works (higher and lower 25\% of the distribution). We observe clear differences between the high and low-quality works of SDG7, which remains constant through all these years. Unfortunately, papers on climate change seem to receive fewer citations than SDG 7, and it seems that SDG 12 will eventually surpass SDG 7. 

\section{Conclusions}

In this work, we have used the tool ASDG \cite{san22} to study the alignment of Aerospace Engineering with the Sustainable Development Goals of the United Nations. ASDG uses NMF, LDA, TopVec, and Bertopic methods to identify the SDGs. The main conclusion of this work is that the introduction of the SDG Agenda did not have a clear impact on the scientific production of the Aerospace Engineering community. This result is quite concerning, as it appears that the community's attention is drifting apart from extraordinary and urgent challenges such as climate change, although differences among countries exist. Funders and research organizations can use ASDG to very easily identify if they are walking on the right path to prosperity. Science maps are possible, even going to smaller organizations like universities. 

One possible limitation of this work is that is based on abstracts, with far better availability than the whole paper. Moreover, at this point, we can only see if there is a connection, but whether this relationship is either positive or negative can not be asserted. So, after this first analysis using ASDG, our objective is to identify the targets of the different SDGs, and annalize their positive and negative interactions. 
Finally, we recommend to any interested reader interested in using the code contact us.

\section*{Acknowledgements}

RV and FFN acknowledge the support of the KTH Climate Action Centre. SHC is partially funded by project PID2021-128676OB-I00 by FEDER/MCIN.

\typeout{} 
\bibliography{ASDG}
\end{document}